\documentclass[1p,final]{elsarticle}
\usepackage{amsfonts,natbib,pslatex}

\usepackage{latexsym,amsmath,amsthm,amssymb,amsxtra,mathrsfs,times,CJK}
\usepackage{makecell}
\usepackage{booktabs}
\usepackage{color}
\usepackage{amsfonts,color}
\usepackage{amssymb,amsmath,latexsym}
\usepackage{amssymb}
\usepackage{extarrows}

\usepackage[colorlinks,linkcolor=blue,anchorcolor=blue,citecolor=blue]{hyperref}
\usepackage[para]{threeparttable}

\newtheorem{theorem}{Theorem}[section]
\newtheorem{lemma}[theorem]{Lemma}

\newtheorem{example}[theorem]{Example}

\journal{Finite Fields and Their Applications}

\begin{document}

\begin{frontmatter}



	\title{Two classes of subfield codes of linear codes}
\author[SWJTU]{Xiaoqiong Ran}
\ead{Ran2021@my.swjtu.edu.cn}

\author[SWJTU]{Rong Luo}
\ead{luorong@swjtu.edu.cn}
 
\cortext[cor1]{Corresponding author}
\address[SWJTU]{School of Mathematics, Southwest Jiaotong University, Chengdu, 610031, China}

\begin{abstract}
	Recently, subfiled codes of linear code over GF$ (q) $ with good parameters were studied, and many optimal subfield codes were obtained. In this paper, Our mainly motivation is to generlize the results of the subfield codes of hyperoval in Ding and Heng (Finite Fields Their Appl. 56, 308-331 (2019)), and generlize the results of two families of subfield codes in Xiang and Yin (Cryptogr. Commun. 13(1), 117-127 (2021)) to $ p $-ary where $ p $ is odd. We get the parameters and weight distribution of these subfield codes. At the same time, the parameters of their dual codes are also studied. When $ m=1 $, The dual codes of these subfield codes are almost MDS code, when $ m>1 $ and $ p $ odd, these dual codes are dimension-optimal with respect to the sphere-backing bound.
\end{abstract}


\end{frontmatter}

%The title of your section 1
\section{Introduction}

Let $ q $ be a prime power. Let GF$ (q) $ be a finite fields with $ q $ elements. An $ [n,k,d] $ linear code over GF$ (q) $ is a $ k$-dimensional subspace of GF$ (q)^n $ with length $ n $ and minimum Hamming distance $ d $, where $ n,k,d $ are positive integer. The dual code of $ C $ is defined by
\[
C^{\perp}=\{y\in{\text{GF}(q)^n: yx^T=0 \text{ for all }x\in{C} }\},
\] 
where $ yx^T $ is the standard inner product of two vectors $ x $ and $ y $. Let $ A_i $ denote the number of codewords with Hamming weight $ i $ of linear code $ C $ over GF$ (q) $. The weight enumerator of $ C $ is defined by $ 1+A_1z+A_2z^2+\cdots +A_nz^n $, and sequence $ (1,A_1,\cdots ,A_n) $ is weight distribution of $ C $. Studying weight distribution of linear code is very important. A linear code $ C $ is called $ t $-weight code if the number of nonzero $ A_i $ in the sequence $ (A_1,A_2,\cdots ,A_n) $ is equal to $ t $. A linear code $ C $ over GF$ (q) $ is said to be dimension-optimal (or distance-optimal ) if there does not exist an $ [n,\ge{k+1},d] $ (or $ [n,k,\ge{d+1}] $) linear code over GF$ (q) $ and almost distance-optimal (or almost dimension-optimal) if $ [n,k,d+1] $ optimal (or $ [n,k+1,d] $) optimal.

Let $ f_i(x_1,x_2,\cdots ,x_t) (1\le{i}\le{a})$ be an $ t$-variable polynomial over GF$ (q) $. Denote
\begin{align*}
G_{x_1,x_2,\cdots ,x_t}=\begin{pmatrix}
f_1(x_1,x_2,\cdots ,x_t)\\
f_2(x_1,x_2,\cdots ,x_t)\\
\cdots \\
f_a(x_1,x_2,\cdots ,x_t)\\
\end{pmatrix}_{(x_1,x_2,\cdots ,x_t)\in{\text{GF}(q)^t}}
\end{align*} 
which is a $ a\times q^t  $ matrix over GF$ (q) $.

Let $ B=[G_{(x_1,x_2,\cdots ,x_t)}|A] $, we define a class of linear codes $ C $ over GF$ (q) $ with generator matrix $ B $, where $ A $ is a $ a\times b $ matrix over GF$ (q) $ consists of $ b $ distinct column vectors of the identity matrix $ I_a $ for $ 0\le{b}\le{a} $.

Recently, Subfield codes of linear code over finite fields with good parameters and optimal were widely studied, which were first considered in \cite{C2000, C1998}, but the authors do not use the name "subfield codes". The notion of subfield codes was first introduced in [\cite{C2013}, p.5117] and a Magma function for subfield codes is actually executed in the Magma package. With the help of Magma, some of optimal or almost optimal subfield codes were obtained.

Heng and Ding obtained some basic properties and results about subfield codes \cite{D2019}, they have studied the subfield codes of the two families of ovoid codes, and the parameters and the weight distributions of the subfield codes of the elliptic quadric codes and Tits ovoid codes were obtained. Meanwhile, Heng and Ding \cite{Z2019, H2022, Q2020, C2020} studied the subfield codes of hyperoval codes, conic codes, maximal arcs and some $ [q+1,2,q] $ MDS codes, \cite{CXiang2021} discussed two families of subfield codes of linear code when $ p=2 $ and obtained some distance-optimal and almost distance-optimal binary codes with a few weights. Wang and Zheng \cite{W2020} extended the work of Heng and Ding in \cite{Z2019}, and studied more subfield codes of linear code. \cite{X2021} studied a class of subfield codes, which are distence-optimal. It can be known from other's research methods, we can construct some subfield codes with good parameters and optimal (or almost optimal) from some know good linear codes. Inspired and motivated by \cite{CXiang2021, Z2019}, we study the subfield codes of linear codes over GF$ (q) $ with $ q $ is a prime power. 

In the rest of this paper. Section 2 introduces some basic properties of characters, character sums, Norms function, Weil sums, subfield codes and Pless power moments over finite fields. Section 3 studies the paremeters of weight distribution of two families of subfield codes from linear codes and their dual codes. The  proofs are presented in Section 4. In section 5, we summarized.

\section{Preliminaries}\label{Section2}

	In this section, we recall the properties of characters, character sums, Weil sums, subsfield codes and the Pless power moments over finite fields which will be needed in later sections.
\subsection{The properties of characters and Weil sums}
Let $ q=p^m $ be a prime power, GF$(q) $ be the finite field with $ q $ elements and $ \alpha $ be a primitive element of GF$ (q) $. Let $ Tr_{q/p} $ denote the trace function from GF$ (q) $ to GF$ (p) $ given by $ Tr_{q/p}(x)=\sum_{i=0}^{m-1}x^{p^i} $ for all $ x\in{\text{GF}(q)} $. Let $ N_{q/p} $ denote the norm function from GF$ (q)^* $ to GF$ (p)^* $, which is represented as
\[
N_{q/p}(x)=x\cdot x^p\cdot x^{p^2}\cdot \cdots\cdot  x^{p^{m-1}}=x^{\frac{p^m-1}{p-1}}\text{ for all }x\in{\text{GF}(q)^*}
\]	

An additive character of GF$ (q) $ is a nonzero function $ \chi$ from GF$(q)$ to $\mathbb{C}^*$, such that
\[
\chi (x+y)=\chi(x)+\chi(y),\text{ for all }x,y\in{\text{GF}(q)},
\]
where $ \mathbb{C}^* $ denotes the set of all nonzero complex numbers. For any $ a\in{\text{GF}(q)} $, the function
\[
\chi_a(x)=\zeta_p^{Tr_{q/p}(ax)},x\in{\text{GF}(q)},
\]	
define an additive character of GF$(q) $, where $ \zeta_p $ is the primitive $ p $-th root of complex unity. If $ a=0, \chi_0(x)=1 $ for all $ x\in{\text{GF}(q)} $, we call it the trivial additive character of GF$(q) $. If $ a=1 $, we refer $ \chi_1 $ as the canonical additive character of GF$(q) $. Obviously, $ \chi_a(x)=\chi_1(ax) $ for all $ x\in{\text{GF}(q)} $. The orthogonality relation of additive characters is given as follows
\begin{align*}
\sum_{x\in{\text{GF}(q)}}\chi_a(x)=\begin{cases}
q&\text{for }a=0,\\
0&\text{for }a\in{\text{GF}(q)^*}.
\end{cases}
\end{align*}

Let GF$(q)^*=\text{GF}(q)\backslash \{0\} $, A multiplicative character of GF$(q)^* $ is a nonzero function $ \psi $ from $ \text{GF}(q)^* $ to $ \mathbb{C}^* $ such that $ \psi(xy)=\psi(x)\psi(y) $ for all $ x,y\in{\text{GF}(q)^*} $. For any $ a\in{\text{GF}(q)^*} $, there is a integer $ k $, such that $ a=\alpha^k $. All the multiplicative characters are given by
\[
\psi_j(\alpha^k)=\zeta_{q-1}^{jk}\text{ for }k=0,\cdots ,q-1,
\]
where $ 0\le{j}\le{q-2} $ and $ \zeta_{q-1}=e^{\frac{2\pi i}{q-1}} $. Furthermore, $ \{\psi_j:j=0,1,\cdots ,q-2 \} $ is multiplicative cyclic group, $ \psi_0 $ is called  the trivial multiplicative character. Let $ \eta:=\psi_{(q-1)/2} $, and is called the quadratic multiplicative character of GF$(q)  $. The orthogonality relation of multiplicative characters is given by
\begin{align*}
\sum_{x\in{\text{GF}(q)^*}}\psi_j(x)=\begin{cases}
q-1&\text{for }j=0,\\
0&\text{for }j\neq{0}.
\end{cases}
\end{align*}

Let $ \chi $ be a additive character of GF$ (q) $, and $ \psi $ be a multiplicative character of GF$ (q)^* $. The Gaussian sum $ G(\chi,\psi) $ is given by
\[
G(\chi,\psi)=\sum_{x\in{\text{GF}(q)^*}}\psi(x)\chi(x),
\]
where $ G(\eta,\chi) $ is called the quadratic sum over GF$ (q) $ and $ \chi $ is nontrivial \cite{L1997}.

\begin{lemma}[\cite{L1997}, Th. 5.15]
	Let $ q=p^m $ with $ p $ odd. $ \chi $ be the canonical additive character of GF$ (q) $. Then
	\begin{align*}
	G(\chi,\psi)&=(-1)^{m-1}(\sqrt{-1})^{(\frac{p-1}{2})^2m}\sqrt{q}\\
	&=\begin{cases}
	(-1)^{m-1}\sqrt{q}&\text{for }p\equiv 1\text{ (mod 4)},\\
	(-1)^{m-1}(\sqrt{-1})^m\sqrt{q}&\text{for }p\equiv 3\text{ (mod 4)}.
	\end{cases}
	\end{align*}
\end{lemma}
Let $ \chi $ be a canonical additive character of GF$ (q) $, and $ f(x) $ is a polynomial over GF$(q)  $ with positive degree. The character sums is given by
\[
\sum_{x\in{\text{GF}(q)}}\chi(f(x))
\] 
is called Weil sums. If $ q $ is odd, the Weil sums is described in the following lemma. 
\begin{lemma}[\cite{L1997}, Th. 5.33] 
	Let $ \chi $ be a nontrivial additive character of GF$ (q) $ with $ q $ odd, and Let $ f(x)=a_2x^2+a_1x+a_0\in{\text{GF}(q)[x]} $ with $ a_2\neq{0} $. Then
	\[
	\sum_{x\in{F_q}}\chi(f(x))=\chi(a_0-a_1^2(4a_2)^{-1})\eta(a_2)G(\eta,\chi).
	\]	
\end{lemma}

\subsection{Pless power moments}
Let $ C $ is an $ [n,k,d] $ linear code over GF$(p)  $ with the weight distribution $ (1,A_1,\cdots ,A_n) $, and $ A_i $ be the number of codeword of weight $ i $ for all $ i=0,1,\cdots ,n $, the weight distribution of dual code $ C^{\perp} $ is denoted by $ (1,A_1^{\perp},\cdots ,A_n^{\perp}) $. Then the first four Pless power moments are given as follows \cite{H2003}:
\begin{align*}
\sum_{i=0}^{n}A_i&=p^k;\\
\sum_{i=0}^{n}iA_i&=p^{k-1}(pn-n-A_1^{\perp});\\
\sum_{i=0}^{n}i^2A_i&=p^{k-2}[(p-1)n(pn-n+1)-(2pn-p-2n+2)A_1^{\perp}+2A_2^{\perp}  ];\\
\sum_{i=0}^{n}i^3A_i&=p^{k-3}[(p-1)n(p^2n^2-2pn^2+3pn-p+n^2-3n+2)-(3p^2n^2-3p^2n-6pn^2+12pn\\
&+p^2-6p+3n^2-9n+6)A_1^{\perp}+6(pn-p-n+2)A_2^{\perp}-6A_3^{\perp}].
\end{align*}

The following lemma give sphere packing bound.
\begin{lemma}[\cite{H2003}, Th. 1.12.1] 
	(Sphere Packing bound) Let $ C $ be a $ q $-ary $ [n,k,d] $ code. Then 
	\[
	q^n\ge{q^k\sum_{i=0}^{\lfloor \frac{d-1}{2} \rfloor}}(q-1)^i\binom{n}{i}
	\]	
\end{lemma}

\subsection{The subfield code from linear code over GF$ (q) $}
Let GF$ (q)^m $ be a finite fields with $ q^m $ elements. Let $ C $ be an $ [n,k,d] $ linear code over  GF$ (q)^m $ with generator matrix $ G $, we can construct a new $ [n,k^{'},d^{'}] $ linear code  over GF$ (q) $ by the following method.

Take a basis of GF$ (q)^m $ over GF$ (q) $. Represent each entry of $ G $ as an $ m\times 1 $ column vector of GF$ (q)^m $ with respect to this basis, and replace each entry of $ G $ with the corresponding $ m\times 1 $ column vector of GF$ (q)^m $. In this way, $ G $ is modified into a $ km\times n $ matrix over GF$ (q) $, which generates the new subfield code $ C^{(q)} $ over GF$(q)  $ with length $ n $. The subfield codes $ C^{(q)} $ is independent of both the choice of the basis of GF$ (q)^m $ over GF$ (q) $ and the choice of the generator matrix $ G $ of $ C $ (see Theorems 2.1 and 2.6 in \cite{D2019}). 

The following two theorems give the trace representation of linear code $ C $, which is called subfield code and a relation between the minimal distance of $ C^{\perp} $ and the minimal distance of $ C^{(p)\perp} $, respectively.
\begin{theorem}[\cite{D2019}, Th. 2.5] 
	Let $ C $  be an $ [n,k] $ linear code over GF$ (q) $. Let $ G=[g_{ij}]_{1\le{i}\le{k},1\le{j}\le{n} } $ be a generator matrix of $ C $. Then the trace representation of the subfield code $ C^{(p)} $ is given by
	\[
	C^{(p)}=\{(Tr_{q/p}(\sum_{i=1}^ka_ig_{i1}) ,\cdots ,Tr_{q/p}(\sum_{i=1}^ka_ig_{in})):a_1,\cdots ,a_k\in{\text{GF}(q)}   \},
	\]
\end{theorem}
\begin{theorem}
	Let $ C $ be an $ [n,k] $ linear code over GF$ (q) $. Then the minimal distance $ d^{\perp} $ of $ C^{\perp} $ and the minimal distance $ d^{(p)\perp} $ of $ C^{(p)\perp} $ satisfy $ d^{(p)\perp}\ge{d^{\perp}} $.
\end{theorem}

\section{The subfield codes from two families of linear codes}

	Let $ q=p^m $ with $ p $ be a prime, and $ U=\{x^2:x\in{\text{GF}(q)} \} $. Let $ f_{11}(x,y)=x $, $ f_{12}(x,y)=y $, $ f_{13}(x,y)=1 $, $ f_{21}(x,y)=\text{Norm}(x) $, $ f_{22}(x,y)=y $, $ f_{23}(x,y)=1 $ be 2-variable polynomials over GF$(q)  $. Denote
\begin{align*}
G_{x,y}=\begin{pmatrix}
x\\
y\\
1
\end{pmatrix}_{x\in{U},\text{ $ y $}\in{\text{GF}(q)}}\text{ and }\bar{G}_{x,y}=\begin{pmatrix}
\text{Norm}(x)\\
y\\
1
\end{pmatrix}_{x\in{\text{GF}(q)^*},\text{ $ y $}\in{\text{GF}(q)}},
\end{align*}
which are $ 3\times \frac{p^{2m}+p^m}{2} $ matrix and $ 3\times q(q-1) $ matrix over GF$ (q) $, respectively. Let
\begin{align*}
G_1=\begin{pmatrix}
z & 1 &0\\
y&0&1\\
1&0&0
\end{pmatrix}_{x\in{U},\text{ $ y $} \in{\text{GF}(q)}} \text{ and }G_2=\begin{pmatrix}
\text{Norm}(x) &1& 0\\
y&0&1\\
1&0&0
\end{pmatrix}_{x\in{\text{GF}(q)^*},\text{ $ y $}\in{\text{GF}(q)}}.
\end{align*}

Two types of linear codes $ C_1 $ and $ C_2 $ over GF$ (q) $ are defined, where their generator matrix are $ G_1 $ and $ G_2 $, respectively.

The subfield codes of $ C_1 $ and $ C_2 $ are described as $ C_1^{(p)} $ and $ C_2^{(p)} $, respectively. By Theorem 2.4, the trace representations are given by
\begin{align*}
C_1^{(p)}=&\{((Tr_{q/p}(ax+by)+c )_{x\in{U},y\in{\text{GF}(q)}},Tr_{q/p}(a),Tr_{q/p}(b) ):(a,b)\in{\text{GF}(q)^2},c\in{\text{GF}(p)}  \}\\
\text{and}\\
C_2^{(p)}=&\{((Tr_{q/p}(a\text{Norm}(x)+by)+c )_{x\in{\text{GF}(q)^*},y\in{\text{GF}(q)}},Tr_{q/p}(a),
Tr_{q/p}(b) ):a,b\in{\text{GF}(q)},\\&c\in{\text{GF}(p)}  \}
\end{align*}

In the following content, our goal is to investigate the parameters and weight distribution of subfield codes $ C_1^{(p)} $ with $ p $ odd, and the parameters and weight distribution of $ C_2^{(p)} $ with $ p $ is prime and $ m=2 $. The following lemma will be used later.
\begin{lemma}\cite{Z2019} 
	Let $ q=p^m $ be an odd prime power. Then the following hold.
	\begin{align*}
	\sharp \{&a\in{\text{GF}(q)^*}:\eta(a)=1\text{ and }Tr_{q/p}(a)=0  \}=\begin{cases}
	\frac{p^{m-1}-1-(p-1)p^{\frac{m-2}{2}}(-1)^{\frac{(p-1)m}{4}} }{2}&\text{ for even }m,\\
	\frac{p^{m-1}-1}{2}&\text{ for odd }m.
	\end{cases}\\
	\sharp \{&a\in{\text{GF}(q)^*}:\eta(a)=1\text{ and }Tr_{q/p}(a)\neq{0}  \}=\begin{cases}
	\frac{(p-1)(p^{m-1}+p^{\frac{m-2}{2}}(-1)^{\frac{(p-1)m}{4}}) }{2}&\text{ for even }m,\\
	\frac{p^{m-1}(p-1) }{2}&\text{ for odd }m.
	\end{cases}	\\
	\sharp \{&a\in{\text{GF}(q)^*}:\eta(a)=-1\text{ and }Tr_{q/p}(a)=0  \}=\begin{cases}
	\frac{p^{m-1}-1+(p-1)p^{\frac{m-2}{2}}(-1)^{\frac{(p-1)m}{4}} }{2}&\text{ for even }m,\\
	\frac{p^{m-1}-1}{2}&\text{ for odd }m.
	\end{cases}\\
	\sharp \{&a\in{\text{GF}(q)^*}:\eta(a)=-1\text{ and }Tr_{q/p}(a)\neq{0}  \}=\begin{cases}
	\frac{(p-1)(p^{m-1}-p^{\frac{m-2}{2}}(-1)^{\frac{(p-1)m}{4}}) }{2}&\text{ for even }m,\\
	\frac{p^{m-1}(p-1) }{2}&\text{ for odd }m.
	\end{cases}	
	\end{align*}
\end{lemma}
The following two theorems is the main results of this paper, which give the parameters and weight distributions of subfield codes $ C_1^{(p)} $ and $ C_2^{(p)} $, respectively, the proofs will be presented in Section 4.
	\begin{theorem}
	Let $ q=p^m $ with $ p $ odd and $ m $ be a positive integer. We have the following results.\\
	\textbf{Table 1} The weight distribution of $ C_1^{(p)} $ with $ m $ odd.
	\begin{center}
		\begin{tabular}{ll}
			\toprule %[2pt]
			Weight & Multiplicity\\
			\midrule
			0& 1\\
			$ \frac{p^{2m}+p^m}{2} $ & $ p-1 $\\
			$ \frac{p^{2m}+p^m-p^{2m-1}-p^{m-1} }{2} $ & $ (p^{m-1}-1)p^m $\\
			$ \frac{p^{2m}+p^m-p^{2m-1}-p^{m-1} }{2}+1 $ & $ (p^m-p^{m-1})(2p^m-p) $\\
			$ \frac{p^{2m}+p^m-p^{2m-1}-p^{m-1} }{2}+2 $ & $ p(p^m-p^{m-1})^2 $\\
			$ \frac{p^{2m}-p^{2m-1} }{2} $ & $ p^{m-1}-1 $\\
			$ \frac{p^{2m}-p^{2m-1} }{2}+1  $ & $ p^m-p^{m-1} $\\
			$ \frac{p^{2m}+p^m-p^{2m-1}-p^{\frac{3m-1}{2}}(-1)^{\frac{(m+1)(p-1)}{4}}}{2} $ & $ \frac{(p^{m-1}-1)(p-1)}{2} $\\
			$ \frac{p^{2m}+p^m-p^{2m-1}-p^{\frac{3m-1}{2}}(-1)^{\frac{(m+1)(p-1)}{4}}}{2}+1 $ & $ \frac{p^{m-1}(p-1)^2}{2} $\\
			$ \frac{p^{2m}+p^m-p^{2m-1}+p^{\frac{3m-1}{2}}(-1)^{\frac{(m+1)(p-1)}{4}}}{2} $ & $ \frac{(p^{m-1}-1)(p-1)}{2} $\\
			$ \frac{p^{2m}+p^m-p^{2m-1}+p^{\frac{3m-1}{2}}(-1)^{\frac{(m+1)(p-1)}{4}}}{2}+1 $ & $ \frac{p^{m-1}(p-1)^2}{2} $\\
			\bottomrule	
		\end{tabular}
	\end{center}
	\textbf{Table 2} The weight distribution of $ C_1^{(p)} $ with $ m $ even.
	\begin{center}
		\begin{tabular}{ll}
			\toprule %[2pt]
			Weight & Multiplicity\\
			\midrule
			0& 1\\
			$ \frac{p^{2m}+p^m}{2} $ & $ p-1 $\\
			$ \frac{p^{2m}+p^m-p^{2m-1}-p^{m-1} }{2} $ & $ (p^{m-1}-1)p^m $\\
			$ \frac{p^{2m}+p^m-p^{2m-1}-p^{m-1} }{2}+1 $ & $ (p^m-p^{m-1})(2p^m-p) $\\
			$ \frac{p^{2m}+p^m-p^{2m-1}-p^{m-1} }{2}+2 $ & $ p(p^m-p^{m-1})^2 $\\
			$ \frac{p^{2m}-p^{2m-1}+(p-1)p^{\frac{3m-2}{2}}(-1)^{\frac{m(p-1)}{4} }}{2} $ & $ \frac{p^{m-1}-1-(p-1)p^{\frac{m-2}{2}}(\sqrt{-1})^{\frac{(p-1)m}{2}}  }{2} $\\
			$ \frac{p^{2m}-p^{2m-1}+(p-1)p^{\frac{3m-2}{2}}(-1)^{\frac{m(p-1)}{4} }}{2}+1 $ & $ \frac{(p-1)(p^{m-1}+p^{\frac{m-2}{2}}(\sqrt{-1})^{\frac{(p-1)m}{2}} ) }{2} $\\
			$ \frac{p^{2m}-p^{2m-1}-(p-1)p^{\frac{3m-2}{2}}(-1)^{\frac{m(p-1)}{4} }}{2} $ & $ \frac{p^{m-1}-1+(p-1)p^{\frac{m-2}{2}}(\sqrt{-1})^{\frac{(p-1)m}{2}}  }{2} $\\
			$ \frac{p^{2m}-p^{2m-1}-(p-1)p^{\frac{3m-2}{2}}(-1)^{\frac{m(p-1)}{4} }}{2}+1 $ & $ \frac{(p-1)(p^{m-1}-p^{\frac{m-2}{2}}(\sqrt{-1})^{\frac{(p-1)m}{2}} ) }{2} $\\
			$ \frac{p^{2m}+p^m-p^{2m-1}-p^{\frac{3m-2}{2}}(-1)^{\frac{m(p-1)}{4} }}{2} $ & $ \frac{(p-1)(p^{m-1}-1-(p-1)p^{\frac{m-2}{2}}(\sqrt{-1})^{\frac{(p-1)m}{2}} ) }{2} $\\
			$ \frac{p^{2m}+p^m-p^{2m-1}-p^{\frac{3m-2}{2}}(-1)^{\frac{m(p-1)}{4} }}{2}+1 $ & $ \frac{(p-1)^2(p^{m-1}+p^{\frac{m-2}{2}}(\sqrt{-1})^{\frac{(p-1)m}{2}} ) }{2} $\\
			$ \frac{p^{2m}+p^m-p^{2m-1}+p^{\frac{3m-2}{2}}(-1)^{\frac{m(p-1)}{4} }}{2} $ & $ \frac{(p-1)(p^{m-1}-1+(p-1)p^{\frac{m-2}{2}}(\sqrt{-1})^{\frac{(p-1)m}{2}} ) }{2} $\\	
			$ \frac{p^{2m}+p^m-p^{2m-1}+p^{\frac{3m-2}{2}}(-1)^{\frac{m(p-1)}{4} }}{2}+1 $ & $ \frac{(p-1)^2(p^{m-1}-p^{\frac{m-2}{2}}(\sqrt{-1})^{\frac{(p-1)m}{2}} ) }{2} $\\	
			\bottomrule	
		\end{tabular}
	\end{center}
	\begin{itemize}
		\item [(1)] If $ m=1 $ and $ p $ odd, then $ C_1^{(p)} $ is $ [\frac{p^2+p}{2}+2,3,\frac{p^2-p+2}{2}] $ code, and $ C_1^{(p)\perp} $ is an almost MDS $ [\frac{p^2+p}{2}+2,\frac{p^2+p}{2}-1,3 ] $ code. If $ p $ is odd and $ m>1 $, then $ C_1^{(p)} $ is a $ [\frac{p^{2m}+p^m}{2}+2,2m+1,\frac{p^{2m}+p^m-p^{2m-1}-p^{\frac{3m-1}{2}} }{2}   ] $ code. For odd $ m $, the weight distribution of $ C_{1}^{(p)} $ is given in Table 1.
		\item [(2)] If $ m\ge{2} $ is even, then $ C_1^{(p)} $ is a $ [\frac{p^{2m}+p^m}{2}+2,2m+1,\frac{p^{2m}-p^{2m-1}-(p-1)p^{\frac{3m-2}{2}}}{2} ] $ code, and its weight distribution is given in Table 2.
		\item [(3)] For any positive integer $ m $, the dual $ C_1^{(p)\perp} $ of $ C_1^{(p)} $ is a dimension-optimal $ [\frac{p^{2m}+p^m}{2}+2,\frac{p^{2m}+p^m}{2} -2m+1,3   ] $ code with respect to the sphere-packing bound and a best known code according to http://www.codetables.de.  
	\end{itemize}
\end{theorem}
	\begin{theorem}
	Let $ q=p^m $ with $ p $ be a prime. Let $ m=2 $. We have the following results.\\
	\textbf{Table 3} The weight distribution of $ C_2^{(p)} $.
	\begin{center}
		\begin{tabular}{ll}
			\toprule %[2pt]
			Weight & Multiplicity\\
			\midrule
			0 & 1\\
			$ p^2(p^2-1) $ & $ p-1 $\\
			$ p^2(p^2-1)+1 $ & $ p-1 $\\
			$ (p^2-p)(p^2-1) $ & $ p(p-1) $\\
			$ (p^2-p)(p^2-1)+1 $ & $ p(p-1)(2p-1) $\\
			$ (p^2-p)(p^2-1)+2 $ & $ (p^2-p)^2 $\\
			$ p^2(p+1)(p-2) +1$ & $ (p-1)^2 $\\
			\bottomrule	
		\end{tabular}
	\end{center}
	\begin{itemize}
		\item [(1)] For any prime $ p $ and $ m=2 $, $ C_2^{(p)} $ is a $ [p^2(p^2-1)+2,4,p^2(p+1)(p-2)+1 ] $ code.
		\item [(2)] If $ p=2 $, then the dual $ C_2^{(2)\perp} $ is a $ [14,10,2] $ almost optimal code according to the tables of best code known maintained at http://www.codetables.de. If $ p\ge{3} $ is an odd, then the dual $ C_2^{(p)} $ is a $ [p^2(p^2-1)+2,p^2(p^2-1)-2,2 ] $ distance-optimal code with respect to the sphere-packing bound.
	\end{itemize}
\end{theorem}
\section{Proofs of the results}
In this section, our main work is to prove two theorems we get in Section 3.

\proof Theorem 3.2. The subfield code of $ C_1^{(p)} $ is given by
\begin{align*}
C_1^{(p)}=\{((Tr_{q/p}(ax+by)+c )_{x\in{U},y\in{\text{GF}(q)}},Tr_{q/p}(a),Tr_{q/p}(b) ):(a,b)\in{\text{GF}(q)^2},c\in{\text{GF}(p)}  \}
\end{align*}

Let $ \chi $ and $ \chi^{'} $ are the canonical additive characters of GF$ (q) $ and GF$ (p) $, respectively. Let $ \eta $ and $ \eta^{'} $ be the quadratic multiplicative characters of GF$ (q)^* $ and GF$ (p)^* $, respectively. For each $ a,b\in{\text{GF}(q) },c\in{\text{GF}(p)} $ and $ p $ odd, denote
\begin{align*}
N_0(a,b,c)&=\sharp\{(x,y):x\in{U},y\in{\text{GF}(q):Tr_{q/p}(ax+by)+c=0}  \}, U=\{ x^2: x\in{\text{GF}(q)}\},\\
N^{'}_0(a,b,c)&=\sharp\{ (x,y)\in{\text{GF}(q)^2}:Tr_{q/p}(ax^2+by)+c=0\},\\
N^{''}_0(a,b,c)&=\sharp\{y:x=0,y\in{\text{GF}(q):Tr_{q/p}(by)+c=0 } \}.
\end{align*}
Through analysis, we conclude that 
\[\tag{1}
N_0(a,b,c)=\frac{N^{'}_0(a,b,c)+N^{''}_0(a,b,c)}{2}.
\]
In order to calculate $ N_0(a,b,c) $, we just need to calculate $ N^{'}_0(a,b,c) $ and $ N^{''}_0(a,b,c) $.

By the orthogonality relation of additive characters,
\begin{align*}
N^{'}_0(a,b,c)&=\frac{1}{p}\sum_{\alpha\in{\text{GF}(p)}}\sum_{(x,y)\in{\text{GF}(p)^2}}\zeta_p^{\alpha (Tr_{q/p}(ax^2+by)+c)}\notag\\
&=p^{2m-1}+\frac{1}{p}\sum_{\alpha\in{\text{GF}(p)^*}}\sum_{(x,y)\in{\text{GF}(q)^2}}\zeta_p^{\alpha (Tr_{q/p}(ax^2+by)+c)}\notag\\
&=p^{2m-1}+\frac{1}{p}\Gamma(a,b,c),\tag{2}
\end{align*}
where $ \Gamma(a,b,c)=\sum_{\alpha\in{\text{GF}(p)^*}}\sum_{(x,y)\in{\text{GF}(q)^2}}\zeta_p^{\alpha (Tr_{q/p}(ax^2+by)+c)} $. We discuss the value of $ \Gamma(a,b,c) $ in the following cases.
\begin{itemize}
	\item [(1)] If $ (a,b,c)=(0,0,0) $. Then $ \Gamma(a,b,c)=p^{2m}(p-1) $.
	\item [(2)] If $ (a,b)=(0,0) $ and $ c\neq{0} $. Then $ \Gamma(a,b,c)=p^{2m}\sum_{\alpha\in{\text{GF}(p)^* }}\zeta_p^{\alpha c}= -p^{2m} $.
	\item [(3)] If $ a=0 $ and $ b\neq{0} $ or $ a\neq{0} $ and $ b\neq{0} $. Then $ \Gamma(a,b,c)=0 $.
	\item [(4)] If $ a\neq{0} $ and $ b=0 $. We have
	\begin{align*}
	\Gamma(a,b,c)&=q\cdot \sum_{\alpha\in{\text{GF}(p)^*}}\zeta_p^{\alpha c}\sum_{x\in{\text{GF}(q)}}\chi(a\alpha x^2)\\
	&=q\cdot\eta (a)G(\eta,\chi)\sum_{\alpha\in{\text{GF}(p)^*}}\zeta_p^{\alpha c}\eta(\alpha)\\
	&=q\cdot\eta(a)G(\eta,\chi)\sum_{\alpha\in{\text{GF}(p)^*}}\chi^{'}(\alpha c)\eta(\alpha).
	\end{align*}
	\begin{itemize}
		\item [(a)] If $ m $ is odd, we have $ \eta(\alpha)=\eta^{'}(\alpha) $ for any $ \alpha\in{\text{GF}(p)^*} $. Then by Lemma 3.1 we have
		\begin{align*}
		\Gamma(a,b,c)&=q\cdot  \eta(a)G(\eta,\chi)\sum_{\alpha\in{\text{GF}(p)^*}}\chi^{'}(\alpha c)\eta^{'}(\alpha)\\
		&=\begin{cases}
		0&\text{  if  } c=0\\
		q\cdot G(\eta,\chi)G(\eta^{'},\chi^{'})\eta(a)\eta^{'}(c) &\text{  if  } c\neq{0}
		\end{cases}\\
		&=\begin{cases}
		0 &\text{  if  }c=0,\\
		q\cdot p^{\frac{m+1}{2}}(-1)^{\frac{(p-1)(m+1)}{4}} &\text{ if  }c\neq{0}, \eta(a)\eta^{'}(c)=1,\\
		q\cdot p^{\frac{m+1}{2}}(-1)^{\frac{(p-1)(m+1)+4}{4}} &\text{ if  }c\neq{0}, \eta(a)\eta^{'}(c)=-1.
		\end{cases}
		\end{align*}
		\item [(b)] If $ m $ is even, then $ \eta(\alpha)=1 $ for any $ \alpha\in{\text{GF}(p)^*} $. Then by Lemma 3.1 we have
		\begin{align*}
		\Gamma(a,b,c)&=q\cdot G(\eta,\chi)\eta(a)\sum_{\alpha\in{\text{GF}(p)^*}}\chi^{'}(\alpha c)\\
		&=\begin{cases}
		q\cdot (p-1)p^{\frac{m}{2}}(-1)^{\frac{m(p-1)+4}{4}} & \text{  if  }c=0, \eta(a)=1,\\
		q\cdot (p-1)p^{\frac{m}{2}}(-1)^{\frac{m(p-1)}{4}} & \text{  if  }c=0, \eta(a)=-1,\\
		q\cdot p^{\frac{m}{2}}(-1)^{\frac{m(p-1)}{4}} & \text{  if  }c\neq{0}, \eta(a)=1,\\
		q\cdot p^{\frac{m}{2}}(-1)^{\frac{m(p-1)+4}{4}} & \text{  if  }c\neq{0}, \eta(a)=-1.
		\end{cases}
		\end{align*}
	\end{itemize}
\end{itemize}
By equation (2) and the discussions above, if $ m $ is odd, we conclude that
\begin{align*}\tag{3}
N^{'}_0(a,b,c)=\begin{cases}
p^{2m} & \text{ for  }(a,b,c)=(0,0,0),\\
0&\text{ for  }(a,b)=(0,0),c\neq{0},\\
p^{2m-1} &\text{ for  }b\neq{0}, \text{ or  } a\neq{0},b=0,c=0,\\
p^{2m-1}+p^{\frac{3m-1}{2}}(-1)^{\frac{(p-1)(m+1)}{4}} & \text{ for  }a\neq{0},b=0,c\neq{0},\eta(a)\eta^{'}(c)=1,\\
p^{2m-1}+p^{\frac{3m-1}{2}}(-1)^{\frac{(p-1)(m+1)+4}{4}} & \text{ for  }a\neq{0},b=0,c\neq{0},\eta(a)\eta^{'}(c)=-1,
\end{cases}
\end{align*}
if $ m $ is even, then
\begin{align*}\tag{4}
N^{'}_0(a,b,c)=\begin{cases}
p^{2m} & \text{ for  }(a,b,c)=(0,0,0),\\
0 &\text{ for  }(a,b)=(0,0),c\neq{0},\\
p^{2m-1} & \text{ for  }a=0,b\neq{0},\text{ or  }a\neq{0},b\neq{0},\\
p^{2m-1}+(p-1)p^{\frac{3m-2}{2}}(-1)^{\frac{m(p-1)+4}{4}} &\text{ for  }a\neq{0},b=0,c=0,\eta(a)=1,\\
p^{2m-1}+(p-1)p^{\frac{3m-2}{2}}(-1)^{\frac{m(p-1)}{4}} &\text{ for  }a\neq{0},b=0,c=0,\eta(a)=-1,\\
p^{2m-1}+p^{\frac{3m-2}{2}}(-1)^{\frac{m(p-1)}{4}} &\text{ for  }a\neq{0},b=0,c\neq{0},\eta(a)=1,\\
p^{2m-1}+p^{\frac{3m-2}{2}}(-1)^{\frac{m(p-1)+4}{4}} &\text{ for  }a\neq{0},b=0,c\neq{0},\eta(a)=-1,
\end{cases}
\end{align*}

By the orthogonality relation of additive characters,
\begin{align*}\tag{5}
N_0^{''}(a,b,c)=\begin{cases}
p^m &\text{ for  }(b,c)=(0,0),\\
0 &\text{ for  }b=0,c\neq{0},\\
p^{m-1} & \text{ for  }b\neq{0},
\end{cases}
\end{align*}
by equation (1), (3)-(5) and discussions above, if $ m $ is odd, we have
\begin{align*}\tag{6}
N_0(a,b,c)=\begin{cases}
\frac{p^{2m}+p^m}{2} & \text{ for  }(a,b,c)=(0,0,0),\\
0 & \text{ for  }(a,b)=(0,0), c\neq{0},\\
\frac{p^{2m-1}+p^{m-1}}{2} &\text{ for  }a=0,b\neq{0},\text{ or  }a\neq{0},b\neq{0},\\
\frac{p^{2m-1}+p^m}{2} & \text{ for  }a\neq{0}, b=0,c=0,\\
\frac{p^{2m-1}+p^{\frac{3m-1}{2}}(-1)^{\frac{(p-1)(m+1)}{4}}   }{2} & \text{ for  }a\neq{0},b=0,c\neq{0},\eta(a)\eta^{'}(c)=1,\\
\frac{p^{2m-1}-p^{\frac{3m-1}{2}}(-1)^{\frac{(p-1)(m+1)}{4}}   }{2} & \text{ for  }a\neq{0},b=0,c\neq{0},\eta(a)\eta^{'}(c)=-1,
\end{cases}
\end{align*}
if $ m $ is even, then
\begin{align*}\tag{7}
N_0(a,b,c)=\begin{cases}
\frac{p^{2m}+p^m}{2} & \text{ for  }(a,b,c)=(0,0,0),\\
0 &\text{ for  }(a,b)=(0,0),c\neq{0},\\
\frac{p^{2m-1}+p^{m-1}}{2} & \text{ for  }a=0,b\neq{0},\text{ or  }a\neq{0},b\neq{0},\\
\frac{p^{2m-1}+p^m-(p-1)p^{\frac{3m-2}{2}}(-1)^{\frac{m(p-1)}{4}}}{2} &\text{ for  }a\neq{0},b=0,c=0,\eta(a)=1,\\
\frac{p^{2m-1}+p^m+(p-1)p^{\frac{3m-2}{2}}(-1)^{\frac{m(p-1)}{4}}}{2} &\text{ for  }a\neq{0},b=0,c=0,\eta(a)=-1,\\
\frac{p^{2m-1}+p^{\frac{3m-2}{2}}(-1)^{\frac{m(p-1)}{4}}}{2} &\text{ for  }a\neq{0},b=0,c\neq{0},\eta(a)=1,\\
\frac{p^{2m-1}-p^{\frac{3m-2}{2}}(-1)^{\frac{m(p-1)}{4}}}{2} &\text{ for  }a\neq{0},b=0,c\neq{0},\eta(a)=-1.
\end{cases}
\end{align*}

For any codeword $ c_1(a,b,c) $ of $ C_1^{(p)} $ with $ m $ is odd, we have
\[
c_1(a,b,c)=((Tr_{q/p}(ax+by)+c)_{x\in{U},y\in{\text{GF}(q)}},Tr_{q/p}(a),Tr_{q/p}(b))
\]
by equation (6) the Hamming distance of codeword $ c_1(a,b,c) $ is given by
\begin{align*}
&wt(c_1(a,b,c))\\
&=\begin{cases}
0 & \text{ for  }(a,b,c)=(0,0,0),\\
\frac{p^{2m}+p^m}{2} & \text{ for  }(a,b)=(0,0),c\neq{0},\\
\frac{p^{2m}+p^m-p^{2m-1}-p^{m-1}}{2} & \text{ for  }a=0,b\neq{0},Tr_{q/p}(b)=0,\\
\frac{p^{2m}+p^m-p^{2m-1}-p^{m-1}}{2}+1 & \text{ for  }a=0,b\neq{0},Tr_{q/p}(b)\neq{0},\\
\frac{p^{2m}+p^m-p^{2m-1}-p^{m-1}}{2} & \text{ for  }a\neq{0},b\neq{0},Tr_{q/p}(a)=Tr_{q/p}(b)=0,\\
\frac{p^{2m}+p^m-p^{2m-1}-p^{m-1}}{2}+1 & \text{ for  }a\neq{0},b\neq{0},Tr_{q/p}(a)=0,Tr_{q/p}(b)\neq{0},\\
\frac{p^{2m}+p^m-p^{2m-1}-p^{m-1}}{2}+1 & \text{ for  }a\neq{0},b\neq{0},Tr_{q/p}(a)\neq{0},Tr_{q/p}(b)=0,\\
\frac{p^{2m}+p^m-p^{2m-1}-p^{m-1}}{2}+2 & \text{ for  }a\neq{0},b\neq{0},Tr_{q/p}(a)\neq{0},Tr_{q/p}(b)\neq{0},\\
\frac{p^{2m}-p^{2m-1}}{2} & \text{ for  }a\neq{0}, b=0,c=0,Tr_{q/p}(a)=0,\\
\frac{p^{2m}-p^{2m-1}}{2}+1 & \text{ for  }a\neq{0}, b=0,c=0,Tr_{q/p}(a)\neq{0},\\
\frac{p^{2m}+p^m-p^{2m-1}-p^{\frac{3m-1}{2}}(-1)^{\frac{(p-1)(m+1)}{4}}  }{2} & \text{ for  }a\neq{0},b=0,c\neq{0},\eta(a)\eta^{'}(c)=1,Tr_{q/p}(a)=0,\\
\frac{p^{2m}+p^m-p^{2m-1}-p^{\frac{3m-1}{2}}(-1)^{\frac{(p-1)(m+1)}{4}}  }{2}+1 & \text{ for  }a\neq{0},b=0,c\neq{0},\eta(a)\eta^{'}(c)=1,Tr_{q/p}(a)\neq{0},\\
\frac{p^{2m}+p^m-p^{2m-1}+p^{\frac{3m-1}{2}}(-1)^{\frac{(p-1)(m+1)}{4}}  }{2} & \text{ for  }a\neq{0},b=0,c\neq{0},\eta(a)\eta^{'}(c)=-1,Tr_{q/p}(a)=0,\\
\frac{p^{2m}+p^m-p^{2m-1}+p^{\frac{3m-1}{2}}(-1)^{\frac{(p-1)(m+1)}{4}}  }{2}+1 & \text{ for  }a\neq{0},b=0,c\neq{0},\eta(a)\eta^{'}(c)=-1,Tr_{q/p}(a)\neq{0},\\
\end{cases}\\
&=\begin{cases}
0 & \text{ with 1 times},\\
\frac{p^{2m}+p^m}{2} & \text{ with $ p-1 $ times},\\
\frac{p^{2m}+p^m-p^{2m-1}-p^{m-1}}{2} & \text{ with $ (p^{m-1}-1)p^m $ times},\\
\frac{p^{2m}+p^m-p^{2m-1}-p^{m-1}}{2}+1 & \text{ with $ (p^m-p^{m-1})(2p^m-p) $ times},\\
\frac{p^{2m}+p^m-p^{2m-1}-p^{m-1}}{2}+2 & \text{ with $ p(p^m-p^{m-1})^2 $ times},\\
\frac{p^{2m}-p^{2m-1}}{2} & \text{ with  $ p^{m-1}-1 $ times},\\
\frac{p^{2m}-p^{2m-1}}{2}+1 & \text{ with  $ p^m-p^{m-1} $ times}\\
\frac{p^{2m}+p^m-p^{2m-1}-p^{\frac{3m-1}{2}}(-1)^{\frac{(p-1)(m+1)}{4}}  }{2} & \text{ with $ \frac{(p^{m-1}-1)(p-1)}{2} $ times},\\
\frac{p^{2m}+p^m-p^{2m-1}-p^{\frac{3m-1}{2}}(-1)^{\frac{(p-1)(m+1)}{4}}  }{2}+1 & \text{ with $ \frac{p^{m-1}(p-1)^2}{2} $ times},\\
\frac{p^{2m}+p^m-p^{2m-1}+p^{\frac{3m-1}{2}}(-1)^{\frac{(p-1)(m+1)}{4}}  }{2} & \text{ with $ \frac{(p^{m-1}-1)(p-1)}{2} $ times},\\
\frac{p^{2m}+p^m-p^{2m-1}+p^{\frac{3m-1}{2}}(-1)^{\frac{(p-1)(m+1)}{4}}  }{2}+1 & \text{ with $ \frac{p^{m-1}(p-1)^2}{2} $ times},
\end{cases}
\end{align*}
when $ (a,b,c) $ runs through GF$(q) \times \text{GF}(q)\times \text{GF}(p) $. Let a linear $ [n,k,d] $ code with weight distribution $ (1,A_1,\cdots ,A_n) $, the weight distribution of its dual is $ (1,A_1^{\perp},\cdots ,A_n^{\perp}) $, where $ A_i $ and $ A_i^{\perp} $ are the number of codewords with weight $ i $, respectively, and $ p^k=\sum_{i=0}^{n}A_i $, $ k $ is denoted as the dimension of $ C_1^{(p)} $, then dimension of $ C_1^{(p)} $ is $ 2m+1 $ as $ A_0=1 $.

When $ m $ is even, we can get similar results and their weight distribution by equation (7) and Lemma 3.1. We omit the details here.

By discussion above, we get the paremeters and weight distribution of $ C_1^{(p)} $, the dual of $ C_1^{(p)\perp} $ has length $ \frac{p^{2m}+p^m}{2}+2 $ and dimension $ \frac{p^{2m}+p^m}{2}-2m+1 $. From the weight distrbution and the first four Pless power moments, we can obtain that $ A_1^{(p)\perp}=A_2^{(p)\perp}=0 $ and $ A_3^{(p)\perp}>0 $. Thus, the minimal distance of $ C_1^{(p)\perp} $ is $ d_1^{(p)\perp}=3 $. By the sphere-packing bound, we have
\begin{align*}
p^{k_1^{(p)\perp}}(\sum_{i=0}^{\lfloor \frac{3-1}{2} \rfloor} (p-1)^i \binom{\frac{p^{2m}+p^m}{2}+2}{i}  )\le{p^{\frac{p^{2m}+p^m}{2}+2}  },
\end{align*}
which implies that $ k_1^{(p)\perp}\le{\frac{p^{2m}+p^m}{2}-2m+1 } $. Thus, the dual codes $ C_1^{(p)\perp} $ are dimension-optimal according to the sphere-packing bound. We have completed the proof of Theorem 3.2.
\qed
\begin{example}
	Let $ p=3 $, $ m $ be a positive integer, $ C_1^{(3)} $ is a the linear code over GF$(3)  $, and is presented in Theorem 3.2.
	\begin{itemize}
		\item [(1)] Let $ m=1 $, $ p=3 $. Then $ C_1^{(3)} $ has parameters $ [8,3,4] $ and weight enumerator
		\[
		1+4x^4+6x^5+14x^6+2x^7,
		\]
		it's almost optimal code according to the tables of best code known maintained at http://www.codetables.de. The dual code $ C_1^{(3)\perp} $ has parameters $ [8,5,3] $, which be an almost MDS code.
		\item [(2)] Let $ m=2 $, $ p=3 $. Then $ C_1^{(3)} $ has parameters $ [47,5,18] $ and weight enumerator
		\[
		1+2x^{18}+2x^{19}+8x^{28}+18x^{30}+90x^{31}+108x^{32}+4x^{36}+8x^{37}+2x^{45},
		\]
		where its dual code has parameters $ [47,42,3] $, which is dimension-optimal with respect to the sphere packing bound.
	\end{itemize}
\end{example}

\proof Theorem 3.3. The subfield code of $ C_2^{(p)} $ is given by
\begin{align*}
C_2^{(p)}=&\{((Tr_{q/p}(a\text{Norm}(x)+by)+c )_{x\in{\text{GF}(q)^*},y\in{\text{GF}(q)}},Tr_{q/p}(a),
Tr_{q/p}(b) ):a,b\in{\text{GF}(q)},\\&c\in{\text{GF}(p)}  \}.
\end{align*}
For each $ a,b\in{\text{GF}(q)} $ and $ c\in{\text{GF}(p)} $ where $ p $ is a prime, denote
\[
N_0(a,b,c)=\sharp\{(x,y):x\in{\text{GF}(q)^*,y\in{\text{GF}(q)}:Tr_{q/p}(a\text{Norm}(x)+by)+c=0  }  \}.
\]

By the orthogonality relation of additive characters,
\begin{align*}
N_0(a,b,c)&=\frac{1}{p}\sum_{\alpha\in{\text{GF}(p)} }\sum_{x\in{\text{GF}(q)^*},y\in{\text{GF}(q)}}\zeta_p^{\alpha ( Tr_{q/p}(a\text{Norm}(x)+by)+c)}\\
&=\frac{q(q-1)}{p}+\frac{q}{p}\sum_{\alpha\in{\text{GF}(p)^*} }\sum_{x\in{\text{GF}(q)^*},y\in{\text{GF}(q)}}\zeta_p^{\alpha ( Tr_{q/p}(a\text{Norm}(x)+by)+c)}\\
&=\frac{q(q-1)}{p}+\frac{1}{p}\Gamma (a,b,c)\tag{8},
\end{align*}
where $ \Gamma(a,b,c)=\sum_{\alpha\in{\text{GF}(p)^*} }\sum_{x\in{\text{GF}(q)^*},y\in{\text{GF}(q)}}\zeta_p^{\alpha ( Tr_{q/p}(a\text{Norm}(x)+by)+c)} $. It seems difficult to calculate the value of exponential sums for any positive integer $ m $, but we can get the value when $ m=2 $.

Let $ m=2 $, we have Norm$(x)=x^{p+1}  $. Next we will discuss the value of $ \Gamma(a,b,c) $.
\begin{itemize}
	\item [(1)] If $ (a,b,c)=(0,0,0) $. Then $ \Gamma(a,b,c)=q(q-1)(p-1) $.
	\item [(2)] If $ (a,b)=(0,0) $ and $ c\neq{0} $. Then $ \Gamma(a,b,c)=-q(q-1) $.
	\item [(3)] If $ a=0 $ and $ b\neq{0} $, or $ a\neq{0} $ and $ b\neq{0} $. Then $ \Gamma(a,b,c)=0 $.
	\item [(4)] If $ a\neq{0}, Tr_{q/p}(a)=0 $ and $ b=0 $. Then
	\begin{align*}
	\Gamma(a,b,c)&=q\sum_{\alpha\in{\text{GF}(p)^*} }\zeta_p^{\alpha c} \sum_{x\in{\text{GF}(q)^*}}\zeta_p^{Tr_{q/p}(a\alpha x^{p+1})}\\
	&=\begin{cases}
	q(q-1)(p-1)&\text{ if  }c=0,\\
	-q(q-1)&\text{ if  }c\neq{0}.
	\end{cases}
	\end{align*}
	\item [(5)] If $ a\neq{0}, Tr_{q/p}(a)\neq{0} $ and $ b=0 $. Then
	\begin{align*}
	\Gamma(a,b,c)=\begin{cases}
	-q(p+1)(p-1)& \text{ if  }c=0,\\
	q(p+1)&\text{ if  }c\neq{0}.
	\end{cases}
	\end{align*}
\end{itemize}
by equation (8) and discussions above, we conclude that
\begin{align*}\tag{9}
N_0(a,b,c)=\begin{cases}
p^2(p^2-1) & \text{ for  }(a,b,c)=(0,0,0), \text{ or  }a\neq{0}, b=c=0,Tr_{q/p}(a)=0\\
0& \text{ for  }(a,b)=(0,0),c\neq{0},\text{ or }a\neq{0},b=Tr_{q/p}(a)=0,c\neq{0}, \\
&\text{ or  }a\neq{0},b=c=0,Tr_{q/p}(a)\neq{0},\\
p(p^2-1) &\text{ for  }b\neq{0},\\
p^2(p+1) & \text{ for  }a\neq{0},b=0,Tr_{q/p}(a)=0,c\neq{0}.
\end{cases}
\end{align*} 

For any codeword $ c_2(a,b,c) $ of $ C_2^{(p)} $. By equation (9), the Hamming weight is given by
\begin{align*}
wt(c_2(a,b,c))&=\begin{cases}
0 & \text{ for }(a,b,c)=(0,0,0),\\
p^2(p^2-1) &\text{ for }(a,b)=(0,0),c\neq{0},\\
(p^2-p)(p^2-1) &\text{ for }a\neq{0},\neq{0},Tr_{q/p}(a)=Tr_{q/p}(b)=0,\\
(p^2-p)(p^2-1)+1 &\text{ for }a\neq{0},\neq{0},Tr_{q/p}(a)=0,Tr_{q/p}(b)\neq{0},\\
(p^2-p)(p^2-1)+1 &\text{ for }a\neq{0},\neq{0},Tr_{q/p}(b)=0,Tr_{q/p}(a)\neq{0},\\
(p^2-p)(p^2-1)+2 &\text{ for }a\neq{0},\neq{0},Tr_{q/p}(a)\neq{0},Tr_{q/p}(b)\neq{0},\\
(p^2-p)(p^2-1) &\text{ for }a=0,b\neq{0},Tr_{q/p}(b)=0,\\
(p^2-p)(p^2-1)+1 & \text{ for }a=0,b\neq{0},Tr_{q/p}(b)\neq{0},\\
0& \text{ for }a\neq{0},b=Tr_{q/p}(a)=c=0,\\
p^2(p^2-1)&\text{ for }a\neq{0},b=Tr_{q/p}(a)=0,c\neq{0},\\
p^2(p^2-1)+1&\text{ for }a\neq{0},b=c=0,Tr_{q/p}(a)\neq{0},\\
p^2(p+1)(p-2)+1&\text{ for }a\neq{0},b=0,Tr_{q/p}(a)\neq{0},c\neq{0}
\end{cases}\\
&=\begin{cases}
0 &\text{ with $ p $ times},\\
p^2(p^2-1) &\text{ with $ p(p-1) $ times},\\
p^2(p^2-1)+1 &\text{ with $ p(p-1) $ times},\\
(p^2-p)(p^2-1) &\text{ with $ p^2(p-1) $ times},\\
(p^2-p)(p^2-1)+1 & \text{ with $ p^2(p-1)(2p-1) $ times},\\
(p^2-p)(p^2-1)+2 &\text{ with $ p(p^2-p)^2 $ times},\\
p^2(p+1)(p-2)+1 &\text{ with $ p(p-1)^2 $ times}.
\end{cases}
\end{align*}
when $ (a,b,c) $ runs through GF$(q)\times \text{GF}(q)\times \text{GF}(p)  $. By observing the calculation above, the Hamming weight 0 occurs $ p $ times. Thus, any codeword of $ C_2^{(p)} $ repeats $ p $ times. Then the dimension of $ C_2^{(p)} $ is 4.

By discussion above, we get the parameters and weight distribution of $ C_2^{(p)} $, the dual is a $ [p^2(p^2-1)+2,p^2(p^2-1)-2] $ code. From the first four Pless power moments, we get $ A_1^{(p)\perp} =0$ and $ A_2^{(p)\perp}>0 $. Thus, the minimal distance $ d_2^{(p)\perp} =2$. By the sphere-packing bound, we have
\begin{align*}
p^{p^2(p^2-1)-2}(\sum_{i=0}^{\lfloor \frac{d_2^{(p)\perp}-1}{2} \rfloor} (p-1)^i\binom{p^2(p^2-1)+2}{i})\le{p^{p^2(p^2-1)+2 }},
\end{align*} 
which implies that $ d_2^{(p)\perp} \le{2}$. Therefore, the dual code $ C_2^{(p)\perp} $ is distance-optimal according to the sphere-packing bound. The proof is completed.
\qed
\begin{example}
	Let $ C_2^{(p)} $ be the linear code over GF$ (p) $, and is presented in Theorem 3.3.
	\begin{itemize}
		\item [(1)] Let $ p=2 $. Then $ C_2^{(2)} $ is a [14, 4, 1] code, the dual is a [14,10,2] best known code according to http://www.codetables.de.
		\item [(2)] Let $ p=3 $. Then $ C_2^{(3)} $ is a [74, 4, 37] code, the dual is a [74,70,2] distance-optimal code with respect to the sphere-packing bound.
	\end{itemize}
\end{example}	
\section{Conclusion}\label{Conclusion}
In this paper, we used the results of Heng and Ding in \cite{Z2019} and continued the work of Xiang and Yin in \cite{CXiang2021}, and mainly investigated the $ p $-ary subfield codes of a class of linear codes. At the same time, the complete weight distribution of $ C^{(p)} $ is presented in Table 1, Table 2 and Table 3. Some linear subfield codes we get in this paper are almost optimal in some cases. The parameters of the dual codes presented in this paper are dimension-optimal (or distance-optimal) according to the sphere-packing bound. From the construction of subfield codes, by selecting generator matrix $ G $ of $ C $, we can get many optimal or almost optimal subfield codes of linear codes $ C $ over GF$ (p) $. It would be interesting to study the weight distributions of subfield codes with good parameters over extended fields.

%\bibliographystyle{AIMS}
%\bibliography{refer}

\end{document}